\documentstyle[9lomcon,cite]{article}

\def\Journal#1#2#3#4{{#1} {\bf #2}, #3 (#4)}

\def\PRL{\em Phys.Rev.Lett.}

\begin{document}
\lctext
\setcounter{page}{1}
\title{Covariant Treatment of Neutrino Spin (Flavour) Conversion in
Matter under the Influence of Electromagnetic Fields}

\author{M.S.Dvornikov, A.M. Egorov, A.E. Lobanov,
A.I. Studenikin\footnote{\normalsize E-mail: studenik@srdlan.npi.msu.su}}

\address{Department of Theoretical Physics, Moscow State University,
119899 Moscow, Russia}


\maketitle\abstracts{Within the recently proposed \cite {ELSt99,
ELStpl00} Lorentz invariant formalism for description of neutrino spin
evolution in presence of an arbitrary electromagnetic fields
effects of matter motion and polarization are considered.}

%\section{}
%\subsection{}

In \cite {ELSt99, ELStpl00}
the Lorentz invariant formalism for neutrino motion in
nonmoving and isotropic matter
under the influence of an arbitrary configuration of electromagnetic fields
have been developed.
We have derived the neutrino spin evolution Hamiltonian that accounts
not only for the transversal to the neutrino momentum components of
electromagnetic field but also for the longitudinal components. With
the using of the proposed Hamiltonian it is possible to consider
neutrino spin precession in an arbitrary configuration of
electromagnetic fields including those that contain strong
longitudinal components. We have also considered the new types
of resonances in the neutrino spin precession
$\nu_{L}\leftrightarrow\nu_{R}$ that could appear when neutrinos propagate
in matter under the influence of different electromagnetic field
configurations.

In the studies \cite {ELSt99,ELStpl00} of the neutrino spin evolution
we have focused mainly on description of influence of different
electromagnetic fields, while modeling the matter we confined
ourselves to the most simple case of nonmoving and unpolarized
matter.  Now we should like go to further and to generalize our
approach for the case of moving and polarized homogeneous matter.

To derive the equation for the neutrino
spin evolution in electromagnetic field $F_{\mu \nu}$ in such a matter we
again start from the
Bargmann-Michel-Telegdi (BMT) equation \cite{BMT59} for the spin vector
$S^{\mu}$ of a neutral particle that has the following form
\begin{equation}
{dS^{\mu} \over d\tau} =2\mu \big\{ F^{\mu\nu}S_{\nu} -u^{\mu}(
u_{\nu}F^{\nu\lambda}S_{\lambda} ) \big\} +2\epsilon \big\{ {\tilde
F}^{\mu\nu}S_{\nu} -u^{\mu}(u_{\nu}{\tilde
F}^{\nu\lambda}S_{\lambda}) \big\},
\label{1}
\end{equation}
This form of the BMT equation corresponds to the case of the particle moving
with constant speed,
$\vec \beta=const$, ($u_\mu=(\gamma,\gamma \vec \beta),
\gamma=(1-\beta^2)^{-1/2})$, in presence of an electromagnetic
field $F_{\mu\nu}$.
Here $\mu$ is the fermion magnetic moment and
${\tilde F}_{\mu\nu}$ is the dual electromagnetic field tensor.
The neutrino spin vector also satisfies the usual conditions,
$S^2=-1$ and $S^{\mu}u_\mu =0$.
Equation (1) covers also the case of a neutral fermion having
static non-vanishing electric dipole moment, $\epsilon$.  Note that
the term proportional to $\epsilon$ violates $T$ invariance.

The BMT spin evolution equation (1) is derived in the frame of
electrodynamics, the model which is $P$ invariant.
Our aim is to generalize this equation for the case when effects of
various neutrino interactions (for example, weak
interaction for which $P$ invariance is broken)
with moving and polarized matter are also taken into account.
Effects of possible $P$ nonconservation and non-trivial properties of matter
(i.e., its motion and polarization) have to be reflected in the equation that
describes the neutrino spin evolution in an electromagnetic field.

The Lorentz invariant
generalization of eq.(1) for our case can be obtained by the substitution
of the electromagnetic field tensor
$F_{\mu\nu}=(\vec E,\vec B)$ in the following way:
\begin{equation}
F_{\mu\nu}\rightarrow F_{\mu\nu}+G_{\mu\nu}.
\label{2}
\end{equation}

In evaluation of the tensor $G_{\mu \nu}$ we demand that the neutrino
evolution
equation has to be linear over
the neutrino spin vector $S_{\mu}$, electromagnetic field $F_{\mu\nu}$,
and such characteristics of matter (which is composed of different fermions,
$f=e,\ n,\ p...$) as fermions currents
\begin{equation}
j_{f}^\mu=(n_f,n_f\vec v_f),
\label{3}
\end{equation}
and fermions polarizations
\begin{equation}
\lambda^{\mu}_f =\Big(n_f (\vec \zeta_f \vec v_f ),
n_f \vec \zeta _f \sqrt{1-v_f^2}+
{{n_f \vec v_f (\vec \zeta_f \vec v_f )} \over {1+\sqrt{1-v_f^2}}}\Big).
\label{4}
\end{equation}
Here $n_f$, $\vec v_f$, and $\vec \zeta_f \
(0\leq |\vec \zeta_f |^2 \leq 1)$ denote, respectively,
the number density of the background fermions f, the
speed of the reference frame in which the mean
momentum of fermions $f$ is zero, and the mean value of the polarization
vectors of the background fermions $f$ in the above mentioned reference frame.
For each type of fermions $f$ there
are only three vectors, $u^{f}_\mu, \ j^{f}_\mu,$ and $\lambda ^{f}_\mu$,
using which the tensor $G_{\mu \nu}$ have to be constructed.
If $j^{f}_\mu$ and $\lambda^{f}_{\mu}$ are slowly varying functions
in space and time
(this condition is similar to one imposed on the electromagnetic
field tensor $F_{\mu \nu}$ in the derivation of the BMT equation)
then one can construct only four tensors (for each of the fermions
$f$) linear in respect to the characteristics of matter:
\begin{equation}
G_{1}^{\mu \nu}=\epsilon ^{\mu \nu \rho \lambda}u_{\lambda}j_{\rho}, \
\ \ G_{2}^{\mu \nu}=\epsilon ^{\mu \nu \rho \lambda}u_{\lambda}\lambda_{\rho},
\label{5}
\end{equation}
\begin{equation}
G_{3}^{\mu \nu}=u^{\mu}j^{\nu}-j^{\mu}u^{\nu}, \ \ \
G_{4}^{\mu \nu}=u^{\mu}\lambda ^{\nu}-\lambda ^{\mu}u^{\nu}.
\label{6}
\end{equation}
Thus,
in general case of neutrino interaction with different background
fermions $f$ we introduce for description of matter effects
antisymmetric tensor
\begin{equation}
G^{\mu \nu}= \epsilon ^{\mu \nu \rho \lambda}
g^{(1)}_{\rho}u_{\lambda}- (g^{(2)\mu}u^\nu-u^\mu g^{(2)\nu}),
\label{7}
\end{equation}
where
\begin{equation}
g^{(1)\mu}=\sum_{f}^{} \rho ^{(1)}_f j_{f}^\mu
+\rho ^{(2)}_f \lambda _{f}^{\mu}, \ \
g^{(2)\mu}=\sum_{f}^{} \xi ^{(1)}_f j_{f}^\mu
+\xi ^{(2)}_f \lambda _{f}^{\mu}.
\label{8}
\end{equation}
Summation is performed over fermions $f$ of the background. The explicit
expressions for the coefficients $\rho_{f}^{(1),(2)}$ and $\xi_{f}^{(1),(2)}$
could be found if the particular
model of neutrino interaction is chosen.
In the usual notations the antisymmetric tensor $G_{\mu \nu}$ can be
written in the form,
\begin{equation}
G_{\mu \nu}= \big(-\vec P,\ \vec M),
\label{9}
\end{equation}
where
\begin{equation}
\vec M= \gamma \big\{(g^{(1)}_0 \vec \beta-\vec g^{(1)})
- [\vec \beta \times \vec g^{(2)}]\big\}, \
\vec P=- \gamma \big\{(g^{(2)}_0 \vec \beta-\vec g^{(2)})
+ [\vec \beta \times \vec g^{(1)}]\big\}.
\label{10}
\end{equation}
It worth to note that the substitution (2) implies that the magnetic $\vec B$
and electric $\vec E$ fields are shifted by the vectors $\vec M$ and $\vec P$,
respectively:
\begin{equation}
\vec B \rightarrow \vec B +\vec M, \ \ \vec E \rightarrow \vec E -\vec P.
\label{11}
\end{equation}

In the case of nonmoving, $\vec v_f=0$, and unpolarized, $\vec \zeta _f=0,$
matter we get, confirming our previous result \cite {ELSt99, ELStpl00},
\begin{equation}
G_{\mu \nu}=\Big(\gamma \vec \beta\sum_{f}^{}\xi^{(1)}_f n_f ,
\gamma \vec \beta\sum_{f}^{}\rho^{(1)}_f n_f \Big).
\label{12}
\end{equation}

We finally
arrive to the following equation for the
evolution of the three-di\-men\-sio\-nal neutrino spin vector $\vec S
$ accounting for the direct neutrino interaction with electromagnetic
field $F_{\mu \nu}$ and matter (which is described by the tensor
$G_{\mu \nu}$):
\begin{equation}
{d\vec S \over dt}={2\mu \over \gamma} \Big[
{\vec S \times ({\vec B_0}+\vec M_0) \Big]+{2\epsilon \over \gamma}
\Big[{\vec S} \times (\vec E_0-\vec P_0)} \Big].
\label{13}
\end{equation}

The derivative in the left-hand
side of eq.(13) is taken with respect to time $t$ in the
laboratory frame, whereas the values $\vec B_0$ and $\vec E_0$ are
the magnetic and electric fields in the neutrino rest frame given in terms
of the transversal, $\vec F_{\perp}$, and longitudinal, $\vec F_{\parallel}$,
in respect to the direction of the
neutrino motion field $\vec F= \ \vec B,\vec E$ components in the
laboratory frame,
\begin{equation}
\begin{array}{c}
\vec B_0=\gamma\big(\vec B_{\perp}
+{1 \over \gamma} \vec B_{\parallel} + \sqrt{1-{1 \over
\gamma^2}}\big[{\vec E_{\perp} \times \vec n}\big]\big),\\
\vec E_0=\gamma\big(\vec E_{\perp} +{1 \over \gamma} \vec E_{\parallel} -
\sqrt{1-{1 \over \gamma^2}}\big[{\vec B_{\perp} \times \vec
n}\big]\big), \vec n={\vec \beta}/\beta.
\label{14}
\end{array}
\end{equation}
The influence of matter on the neutrino spin evolution in eq.(13) is given by
the vectors $\vec M_0$ and $\vec P_0$ which in the rest frame of neutrino
can be expressed in terms of quantities determined in the laboratory
frame
\begin{equation}
\begin{array}{c}
\vec M_0=\gamma \vec \beta
\Big(g^{(1)}_0-{{\vec \beta \vec g^{(1)}} \over {1+\gamma ^{-1}}}\Big)
-\vec g^{(1)},
\end{array}
\label{15}
\end{equation}
\begin{equation}
\vec P_0=-\gamma \vec \beta
\Big(g^{(2)}_0-{{\vec \beta \vec g^{(2)}} \over {1+\gamma ^{-1}}}\Big)
+\vec g^{(2)}.
\label{16}
\end{equation}

Let us determine the coefficients $\rho^{(i)}_f$ and $\xi^{(i)}_f$
in eq.(8) for the particular
case of the electron neutrino propagation in moving and polarized
electron gas. We consider the standard model of interaction supplied with
$SU(2)$-singlet right-handed neutrino $\nu_R$. The neutrino effective
interaction Lagrangian reads
\begin{equation}
L_{eff}=-f^\mu \Big(\bar \nu \gamma_\mu {1+\gamma^5 \over 2} \nu \Big),
\label{17}
\end{equation}
where
\begin{equation}
f^\mu={G_F \over \sqrt2}\Big((1+4\sin^2 \theta _W) j^\mu_e -
\lambda ^\mu _e\Big).
\label{18}
\end{equation}
In this case neutrino electric dipole moment vanishes, $\epsilon
=0$, so that the coefficients $\xi^{(i)}_e =0$, and from the obvious
relation, $f_\mu=2\mu g_\mu^{(1)}$, it follows
\begin{equation}
\rho^{(1)}_e={G_F \over
{2\mu \sqrt2}}(1+4\sin^2 \theta _W), \ \rho^{(2)}_e=-{G_F \over {2\mu
\sqrt2}}.
\label{19}
\end{equation}
If for the neutrino magnetic moment we take the
vacuum one-loop contribution \cite {LeeShr77, Fu}
$$\mu_{\nu}={3
\over {8\sqrt2 \pi^2}}eG_F m_\nu,$$
then
$$\rho^{(1)}={4\pi^2\over 3em_\nu}(1+4\sin^2\theta _W),\ \
\rho^{(2)}=-{4\pi^2\over 3em_\nu}.$$

We should like to note that solutions of the derived eq.(13)
for the neutrino spin evolution in moving and polarized matter and,
correspondingly, the neutrino oscillation probabilities and effective
mixing angles $\theta_{eff}$ can be obtained for different
configurations of electromagnetic fields in a way similar to that
described in \cite{ELSt99, ELStpl00, ELS00}.

Let us consider the case of neutrino propagating in the relativistic
flux of electrons. Using expressions for the vector $\vec M_0$, eqs.
(8), (15), we find,

\begin{equation}
\begin{array}{c}
{\vec {M}_0}=n_e\gamma{\vec\beta}\Big\{
\big(
\rho^{(1)}+\rho^{(2)}{\vec\zeta_{e}}{\vec v}_e
\big)
(
1-{\vec\beta}{\vec v}_e
)+\\
+\rho^{(2)}\sqrt{1-v^2_e}
\Big[
{
(\vec \zeta_{e}{\vec v}_e)(\vec\beta{\vec v}_e)
\over
1+\sqrt{1-v^2_e}
}-\vec\zeta_{e}\vec\beta
\Big]
+O(\gamma^{-1})
\Big\}.
\end{array}
\label{M_0} \end{equation}
In the case of slowly moving matter, $v_e\ll1$, we get
\begin{equation}
{\vec M}_0=n_e\gamma{\vec\beta}\Big(
\rho^{(1)}-\rho^{(2)}\vec\zeta_{e}\vec\beta \Big).
\label{21}
\end{equation}
For the unpolarized matter eq.(\ref{21}) reproduces the
Wolfenstein term and confirms our previous result
\cite{ELSt99,ELStpl00}.  In the opposite case of relativistic flux,
$v_e\sim 1$, we find,
\begin{equation}
{\vec M}_0=n_e\gamma{\vec\beta}
\Big(
\rho^{(1)}+\rho^{(2)}\vec\zeta_{e}{\vec v}_e
\Big)
\Big(
1-\vec\beta{\vec v}_e
\Big).
\end{equation}
If we
introduce the invariant electron number density, \begin{equation}
\begin{array}{c}
n_0=n_e\sqrt{1-v^2_e},
\end{array}
\end{equation}
then it follows,
\begin{equation}
\begin{array}{c}
{\vec M}_0=n_o\gamma{\vec\beta}
{
(
1-\vec\beta{\vec v}_e
)
\over
\sqrt{1-v^2_e}
}
\Big(
\rho^{(1)}+\rho^{(2)}\vec\zeta_{e}{\vec v}_e
\Big).

\end{array}
\end{equation}
Thus in the case of the parallel motion of neutrinos
and electrons of the flux matter effect contribution to the neutrino 
spin evolution equation (\ref{13}) is suppressed. It should be noted 
that this fenomenon can exist also for neutrino flavour oscillations 
in moving matter.

\section*{Acknowledgements} We
should like to thank Anatoly Borisov, Pavel Eminov and Pavel Nikolaev
for discussions on the relativistic quantum statistics.


\section*{References}
\begin{thebibliography}{99}

\bibitem{ELSt99} A.Egorov, A.Lobanov, A.Studenikin, in: New Worlds in
                Astroparticle Physics, ed. by A.Mourao, M.Pimento, P.Sa,
                World Scientific, Singapore, p.153, 1999; hep-ph/9902447.

\bibitem{ELStpl00} A.Egorov, A.Lobanov, A.Studenikin, Phys.Lett.B491 (2000)
                   137, hep-ph/9910476.

\bibitem{BMT59} V.Bargmann, L.Michel, V.Telegdi,
                \Journal{\PRL}{2}{435}{1959}.

\bibitem{LeeShr77} B.Lee, R.Shrock, \Journal{PR}{16}{1444}{1977}.

\bibitem{Fu} K.Fujikawa, R.Shrock, \Journal{\PRL}{45}{963}{1980}.

\bibitem{ELS00} A.Egorov, A.Lobanov, A.Studenikin, in: Results and
                Perspectives in Particle Physics, ed. by M.Greco,
                Frascati Physics Series, 2000.

\end{thebibliography}
\end{document}